Floating Mirrors

Ermanno F. Borra[1], Anna Ritcey[2], and Etienne Artigau[1]

Université Laval, Québec, Qc, Canada G1K 7P4

SUBJECT HEADINGS: telescopes

(SEND ALL EDITORIAL CORRESPONDANCE TO: E.F. BORRA)

RECEIVED________________________________________

1 Centre d'Optique, Photonique et Lasers, Département de Physique borra@phy.ulaval.ca, etienne.artigau@phy.ulaval.ca

2 Département de Chimie, anna.ritcey@chm.ulaval.ca

# ABSTRACT

We discuss a new technology that promises large inexpensive mirrors. We argue that it should be possible to tilt a rotating viscous liquid by perhaps as much as a few tens of degrees. The tilted liquid parabolic surface is used as the support for a thin reflecting metallic film. We demonstrate two critical steps: that a viscous liquid mirror can be tilted and that an optical quality metallic film can be deposited on a liquid. The advent of astronomically useful tilted floating mirror telescopes is contingent on the development of high-viscosity high-reflectivity liquids. It is a good omen that we already have identified two classes of such liquids; however there remain technical challenges to overcome before such liquids can be used in viable telescopes.

## 1. INTRODUCTION

In the late 1970s the astronomical community came to the consensus that there was a need for a new generation of ground-based optical telescopes having substantially larger diameters than the 4-m class reflectors that then dominated ground based optical astronomy. This has led to a new generation of optical telescopes in the 8-m class. However the mirrors still use polished glass and, while cheaper per unit collecting area than the older generation, are still quite expensive. These mirrors although innovative essentially push the glass technology to its limit and it is not clear whether they can be used for the still larger telescopes now envisioned. It is therefore clearly worthwhile to explore the use of totally different technologies that would enable us to build large mirrors at low costs. This is the aim of this article.

Liquid mirror telescopes offer a totally different solution. They use the fact that the surface of a liquid rotating in a gravitational field takes the shape of a parabola to give excellent surface qualities at low costs (Borra 1995). For a given diameter, the cost of a mercury liquid mirror is almost two orders of magnitudes less than the cost of a glass mirror, making it very competitive for some applications. The technology is very young but is well established in the laboratory (Girard & Borra 1997) as well as in observatory settings (Sica et al. 1995, Hickson & Mulrooney 1997, Cabanac , Borra, & Beauchemin 1998). Because of its youth, the technology has significant room for improvement.

Mercury liquid mirrors suffer from a major limitation: They cannot be tilted and can only observe the zenith. This greatly restricts their usefulness. In this article, we suggest a new technology that promises very inexpensive mirrors. We argue that it should be possible to tilt a rotating viscous liquid by perhaps as much as a few tens of degrees. The tilted liquid parabolic surface could then be used as the support for a very thin reflecting film. We demonstrate two critical steps for this new technology: that a viscous liquid mirror can be tilted and that an optical quality metallic film can be deposited on a liquid.

## 2. TILTED LIQUID MIRRORS.

Simple logic indicates that a liquid mirror <u>must</u> be tiltable to some extent. If the mirror cannot be tilted, it must be absolutely vertical: an obvious physical impossibility. Indeed, the diffraction limited wavefronts reported by Borra et al. (1992) and Girard & Borra (1997) are obtained with mirrors that have alignment errors of the order of 0.25 arcseconds. Borra et al. (1992) also show wavefronts of mercury liquid mirrors tilted by a couple of arcseconds that have good surface qualities for most of their surfaces. Published experimental evidence therefore shows that liquid mirrors can be tilted. However, a liquid mirror tilted by a few arcseconds is no more useful than a perfectly vertical mirror; one therefore must find ways to increase its tilt by several degrees. How can we do this? Simple considerations show that one must increase the viscosity of the liquid.

Let us consider a small element of liquid in a rotating container which is tilted by an angle θ with respect to the vertical. At the moment when it is at the highest point of its trajectory it is submitted to a force of magnitude gsinθ directed towards the center of the mirror. Half a period later, it is submitted to a force of same magnitude but directed away from the center of the mirror. In the reference frame of the turntable, a liquid element is submitted to a rotating force vector having the period of rotation of the mirror. If the viscosity is sufficiently large that the liquid has moved less than λ/20 in half a turn, it will never move more than λ/20. This qualitative simple analysis neglects Coriolis forces, resonant effects, viscosity effects, interactions with the air and the bottom of the container, gyroscopic effects, etc. A full hydrodynamic treatment must be done for a proper quantitative analysis.

Modeling the effect of a periodic force on a rotating liquid in a circular container is non-trivial and beyond the scope of this article; but we can gain some insight by examining a simpler problem that has an analytical solution. Lamb (1945) treats the problem of a periodic force X = f cos(ωt) acting uniformly on an infinite mass of liquid of uniform depth h. It can be modeled by the differential equation

$$\frac{\partial u}{\partial t} = \nu \frac{\partial^2 u}{\partial y^2} + X, \qquad (1)$$

where u = dx/dt is the speed in the horizontal direction, y is the vertical direction with y=h at the surface and ν is the kinematic viscosity. With the boundary conditions u = 0 at y = 0 and ∂u/∂y=0 at y =h, the solution in the case of high viscosity and low thickness, which applies to our case, can be approximated by

$$u = \frac{f}{2\nu} y(2h - y).\cos(\omega t + \varepsilon). \qquad (2)$$

Equation 2 indicates that perturbations scale as f/ν so that increasing the viscosity ν is equivalent to decreasing the driving force. In our case f= gsinθ , so that perturbations scale as sinθ /ν.

The wavefront of a rotating liquid mirror tilted by 2 arcseconds has been published by Borra et al. (1992), showing that most of the surface is undisturbed and has high quality. A mirror tilted by 10 degrees, feels a force 40,000 greater than a mirror tilted by 1 arcsecond. If perturbations scale as sinθ /ν , as suggested by Eq. 2, one would need a liquid having a kinematic viscosity 40,000 greater than the kinematic viscosity of mercury to get a reasonable surface quality. We therefore need a liquid having a kinematic viscosity of 4,000 cs, roughly the viscosity of glycerin at 10 ° C.

Since the assumptions leading to Eq. 2 differ from the problem at hand, it is not clear to what extent it applies to our problem. There is the obvious difference due to rotation. Furthermore, because our container is not infinite, there are different boundary conditions at the center and at the periphery. A thorough theoretical analysis as well as optical tests are needed. Because we have the laboratory equipment at hand, it is far easier for us to do laboratory work, which should be done along with theoretical work anyway.

## 3 . OPTICAL TESTS OF TILTED ROTATING LIQUID SURFACES

We have carried out exploratory experimental work to determine whether the ideas put forth in this paper are valid. To gain insight into the dynamics of a tilted rotating liquid surface, we have tested liquid mirrors that use silicon oils having different viscosities.

We use the same setups utilized with mercury liquid mirrors (Borra et al. 1992). Optical testing is done with Ronchi tests (Cornejo-Rodriguez 1978) in a small testing tower used to test gallium liquid mirrors as described in Borra, Tremblay, Huot & Gauvin (1997).

The Ronchi test measures slopes and, like the knife-edge test to which it is related, it can be quite sensitive to slope errors and is capable of detecting defects smaller than 1/100 wave. The bars of our Ronchi ruling subtend 12 arcseconds so that slope deviations less than 1 arcsecond can easily be seen. We do not use null lenses; hence the Ronchigrams should show the typical spherically aberrated signatures of parabolas tested at their centers of curvature (Cornejo-Rodriguez 1978). Because the test is qualitative, and because we are only interested in tilt effects, we qualitatively compare the Ronchigrams of the tilted mirrors to the Ronchigrams of the vertical mirrors.

Figure 1 shows Ronchigrams of a 1-m diameter rotating container filled with silicone oil having a viscosity of 1000 cs and average thickness of 1.6 millimeters. They have been taken at tilt angles varying from the vertical to 1.1 degrees. Figure 1 shows that the large scale appearances of the Ronchigrams of the tilted mirrors are qualitatively similar to those of the vertical mirror, suggesting that the large scale figure of the mirror does not degrade much with tilt angle. The shapes of the Ronchigrams in Fig. 1 are not perfectly identical but this is to be expected since their shapes are expected to depend on the position of the ruling on the caustic, as illustrated in figure 9.9 of Cornejo-Rodriguez (1978). Note however, that our qualitative tests do not prove that the surface remains parabolical as it is tilted, they simply indicate that its overall shape remains aspherical and is compatible with the shape of a parabola. On the other hand, should they occur, small large-scale deviations from the ideal shape are not worrisome since they can be corrected with active optics.

As the tilt angles increase, the Ronchigrams become marred by an increasing number of speckles. Their dimensions do not change noticeably with tilt angle. The speckles rotate with the mirror and are stable in position over a few minutes but do not reappear at the same locations after we change a parameter (e.g. tilt angle). Fig. 2 shows the Ronchigram of a container with 500 cs oil 1.6-mm deep tilted 32 arcminutes, as well as a container with 1000 cs oil 3.2-mm deep tilted 65 arcminutes. Comparing Fig. 2 to Fig.1, we can see that the number of speckles decreases with increasing viscosity and with decreasing thickness.

The speckling phenomenon is unexpected and is something we must investigate but it is not particularly worrisome since the number of speckles depends on thickness and viscosity and therefore can be controlled. Speckles may be caused by triboelectricity (friction induced charges). Triboelectricity was expected on mercury mirrors and we have occasionally seen speckles on mercury mirror similar to those shown in figures 1 and 2; although , with mercury, the speckles last far less, presumably because of the lower viscosity.

## 4.        DEPOSITION OF REFLECTIVE FILMS ON LIQUIDS

One of us has been aware for a long time that liquid mirrors can be tilted. However, until now, this knowledge could not be put to use in the absence of a credible candidate high-viscosity, high reflectivity liquid. The situation has now changed.

Tilted liquid mirrors can be useful only if one can make them with high-viscosity high-reflectivity liquids. This can be done by depositing a high-reflectivity metallic layer on a viscous liquid. The underlying liquid then merely serves as a continuous support system for an ultra-thin mirror. We have identified two different techniques that allow one to deposit a high-reflectivity layer on a liquid.

The first one is of our own creation. Our general approach involves the selective deposition of a thin metal layer on an organic polymer film spread at the liquid-air interface. The process relies on the reduction of metal ions in solution by organic molecules that are located only at the surface.  Preliminary results are promising. Interferometric tests on small samples have shown interference fringes. However, we have just begun exploratory experiments and much work remains to be done.

The second technique was found in the chemical literature. Novel silver liquidlike interfacial films were first reported by Yogev and Efrima (1988).  These Metal Liquid-Like

Films (MELLFs ) exhibit the optical properties of metallic silver and, at the same time, striking fluid behaviour. The films flow like liquids and heal rapidly upon rupture. They are quite versatile and can be made with a variety of techniques and chemicals, allowing one to fine-tune their characteristics.

MELLFs are typically formed at the interface between an aqueous phase and a dense organic solvent, such as dichloromethane. Silver is reduced in the aqueous phase, in the presence of certain additives, including surfactants. As the metallic particles form, surfactant molecules adsorb on their surface, inducing particle aggregation. The heavy aggregates fall to the interface, where they are redispersed by the solubility of the hydrophobic surfactant chains in the organic phase. The silver particles then float at the water-organic interface as a stable colloidal suspension. Another technique involves vigorous shaking of a mixture of colloidal particles and an acqueous solution (Vlckova et al. 1993). This last technique seems particularly interesting since it gives silver grains having diameters of the order of a few nanometers (Srnova et al. 1998), insuring low scattered light. The reflecting film so created can be collected with a pipet and then spread over the surface of an acqueous solution. MELLFs prepared with a shaking technique have reflectivities of the order of 80% (Efrima, private communication).

Interferometric tests of samples show that it is possible to generate optical quality surfaces as can be seen in Figure 3 that shows interferometric fringes of a 5 cm wide sample of one of our MELFFs. There are several technical hurdles that must be overcome before we can use them in a tilted liquid mirror. However, we believe that we can overcome them. For example, although the liquids presently made have low viscosities, we know methods that can be used to increase them.

**5. DISCUSSION AND CONCLUSION**

We have suggested a new technique to make parabolic mirrors. It rests on the assumption that it should be possible to tilt by as much as a few tens of degrees the parabola generated by a rotating viscous liquid and use it as the support for a thin reflecting film. In this article we independently demonstrate the two critical steps for this new technology: that a viscous liquid mirror can be tilted and that an optical quality film can be deposited on a liquid.

Ronchi tests show that rotating containers tilted by as much as one degree and covered with silicone oil retain shapes consistent with those of tilted parabolas. The results are consistent with the prediction of Eq. 2: that the maximum tilt angle increases linearly with viscosity. This can be seen from the fact that a naïve extrapolation from the 2 arcsecond tilted mercury mirror (Borra et a. 1992) predicts a good figure at 2.5 degrees for a viscosity of 1000 cs and 1.25 degrees at 500 cs: A prediction consistent with the results shown in Figure 1. We should thus expect that a tilt angle of 20 degrees should be achieved with a liquid having a viscosity of the order of a few $10^4$ centistokes.

The advent of astronomically useful tilted floating mirror telescopes is contingent on the development of high-viscosity high-reflectivity liquids. We have identified two techniques that can deposit a reflecting metallic film on a liquid, one of our own creation and the other from the chemical literature. Metal Liquid Like Films (MELFFs) have been discovered a decade ago and their technology is more developed than ours. The information gathered in the literature and in private conversations with experts on MELFFs indicates that they are very good candidates for our purpose. They are easy to make, inexpensive, robust, versatile, and seem to have the appropriate physical properties. Our interferometric tests on small samples indicate that they can have good surface qualities.

We have independently demonstrated the two main pieces of the puzzle and now we must put them together and produce an astronomically useful mirror. This will take some time and effort. The main challenge is to deposit a high-reflectivity layer on a high-viscosity liquid. We have candidate techniques to do this.

It is a good omen that we already have identified two candidate liquids. One must do additional research to find other classes of liquids and improve the characteristics of the present ones. Several challenges must be met before the technology can be used in astronomical telescope. For example, the viscous liquid should have a freezing temperature below –10 degrees C. Also, the viscous liquids needed to tilt a mirror by several degrees take a long time to stabilize after startup. One must find ways to have it at a lower viscosity at startup and increase its viscosity after stabilization. Fortunately, the viscosity of a liquid decreases with increasing temperature, giving us an obvious solution. Spraying the warm liquid uniformly over the container will also help.

How far should we tilt a mirror to significantly increase its usefulness? Consider that a zenith liquid mirror telescope has a field of regard given by its corrector that is of the order of 1 degree. Tilting it by one degree, increases its field of regard to 3 degrees, tripling it. This is already a significant improvement, especially considering that the use of a low-density liquid substantially decreases costs with respect to a heavy mercury mirror. The cost of a liquid mirror is dominated by the costs of bearing and container and the costs of both depend on the weight of the mirror. However, the real payoff will come if one can tilt the mirror by ten degrees or more and track in real time by changing the tilt. Consider that an astronomical telescope seldom observes further away than 45 degrees from the zenith. If we tilt a mirror by 20 degrees, it accesses half the region of sky accessible to a conventional telescope. If tiltable floating mirror telescopes cost a small fraction of the cost of conventional telescopes, as seems reasonable, they become highly competitive with respect to conventional telescopes. Note that the savings will not only come from the mirror itself but also from the mount since the mirror should be considerably lighter than a glass mirror. The mirror is a gyroscope that resists tilt changes so that the tilt changes must be slow. It is unlikely that the telescope will slew as fast as a conventional telescope.

Although the first reaction is to call this a liquid mirror, it really is a hybrid between a liquid mirror and a solid metal-coated mirror. In a classical glass mirror, the only function of the glass is really to provide an accurate and stable support to a thin coat of reflective metal. The uncoated glass is therefore just as much a part of the support system as the mechanical support system of the glass. In our case, the viscous liquid serves out the purpose of the glass and its support. The beauty of our system is that, while the basic laws of Physics conspire to deform the parabolic shape of a coated glass mirror, they conspire to keep the parabolic shape of our mirror.

Finally, we should point out that the continuous liquid support given by the rotating liquid could be used to support a conventional very thin glass (or other solid) metal coated mirror. Considerable savings will be realized with the simplicity of the support system and the weight reduction. As a bonus, the centrifugal force caused by the rotation of the mirror induces a steady downflow of air that flushes the mirror's surfaces and to bring it to thermal equilibrium with the ambient air.

It is our hope that this article will stimulate further research, particularly on the development of reflecting liquids. The impact on astronomy of low cost large tiltable floating mirror telescopes will be such that it must be done.

**AKNOWLEDGEMENTS.**

We wish to thank Dr. S. Efrima for numerous discussion and Dr. B. Vlckova for numerous discussions and allowing us to visit her laboratory. This research has been supported by the Natural Sciences and Engineering Research Council of Canada

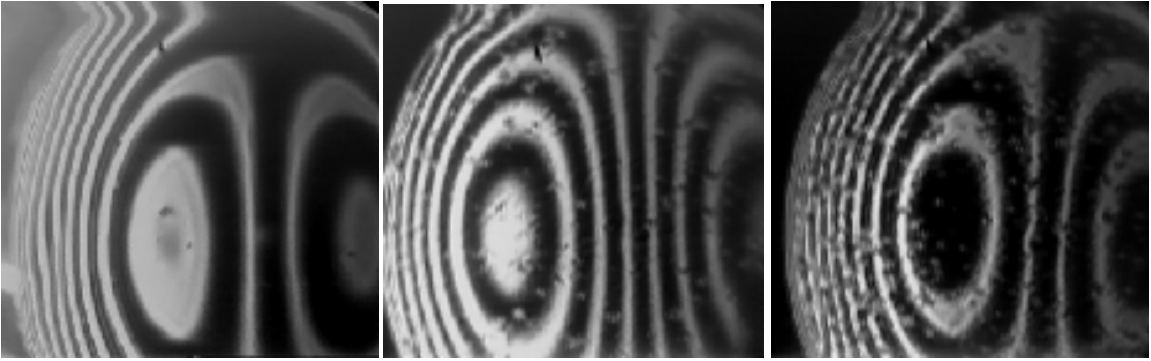

Figure 1: It shows Ronchigrams of a 1-m diameter rotating container filled with silicone oil having a viscosity of 1000 centistokes and average thickness of 1.6 millimeters. They have been taken at tilt angles of 0 degrees (left), 32 arcminutes (center) and 1.1 degrees (right0.

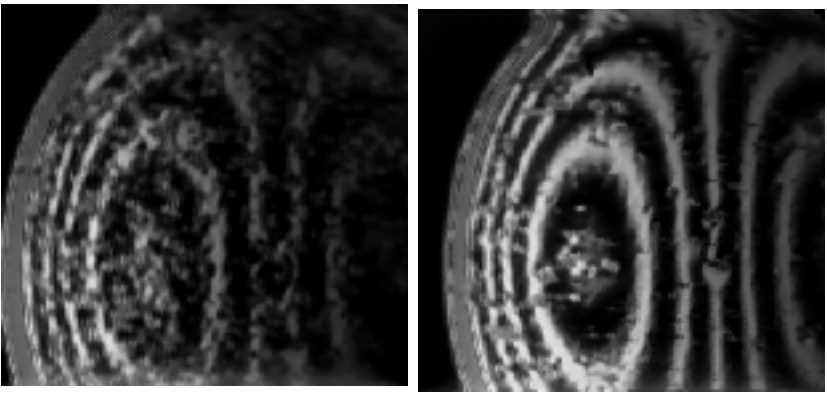

Figure 2:   It shows the Ronchigram of a container filled with 1000 centistokes oil 3.2-mm deep tilted 65  arcminutes (left), as well as a container filled with 500 centistokes oil 1.6-mm deep tilted 32 arcminutes (right).

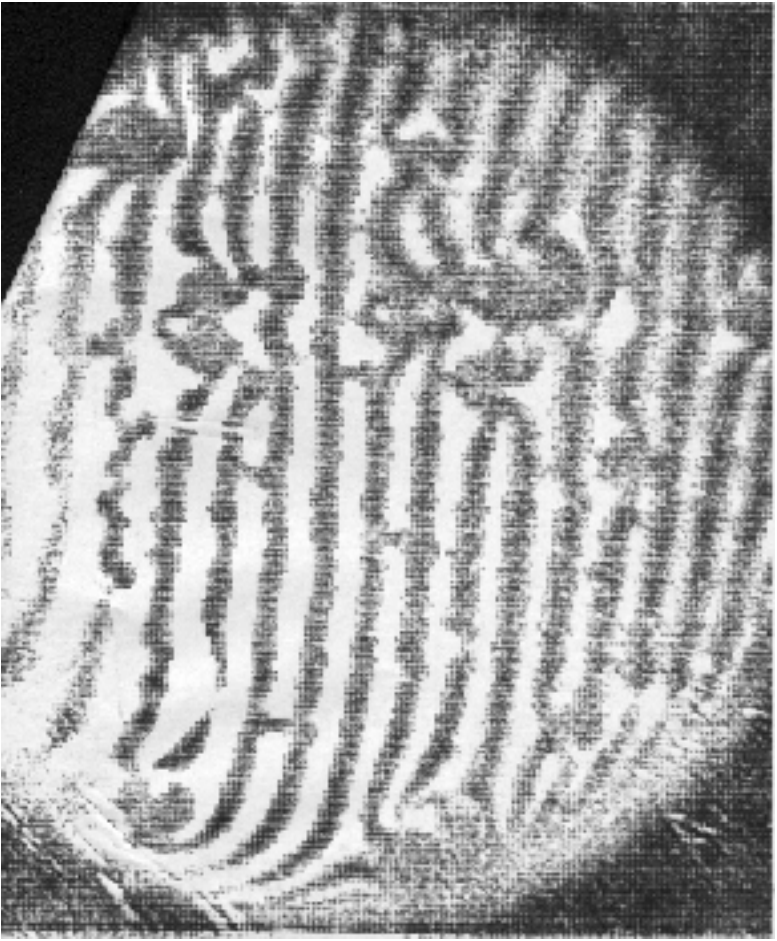

Figure 3: It shows interferometric fringes of a 5 cm wide sample of a MELFF surface.